# Coherent quantum phase slip

O. V. Astafiev[1,2], L. B. Ioffe[3], S. Kafanov[1,2], Yu. A. Pashkin[1,2,4], K. Yu. Arutyunov[5,6], D. Shahar[7], O. Cohen[7], & J. S. Tsai[1,2]

[1]*NEC Green Innovation Research Laboratories, 34 Miyukigaoka, Tsukuba, Ibaraki, 305-8501, Japan*

[2]*The Institute of Physical and Chemical Research (RIKEN), 34 Miyukigaoka, Tsukuba, Ibaraki, 305-8501, Japan*

[3]*Center for Materials Theory, Department of Physics and Astronomy, Rutgers University, 136 Frelinghuysen Rd, Piscataway, NJ 08854, USA*

[4]*Department of Physics, Lancaster University, Lancaster, LA1 4YB, UK*

[5]*University of Jyväskylä, Department of Physics, PB 35, 40014 Jyväskylä, Finland*

[6]*Moscow State University, Institute of Nuclear Physics, Leninskie gory, GSP-1, Moscow 119899, Russia*

[7]*Department of Condensed-Matter Physics, Weizmann Institute of Science, Rehovot 76100, Israel*

**A hundred years after discovery of superconductivity, one fundamental prediction of the theory, the coherent quantum phase slip (CQPS), has not been observed. CQPS is a phenomenon exactly dual**[1] **to the Josephson effect: whilst the latter is a coherent transfer of charges between superconducting contacts**[2,3]**, the former is a**

**coherent transfer of vortices or fluxes across a superconducting wire. In contrast to previously reported observations[4,5,6,7,8] of incoherent phase slip, the CQPS has been only a subject of theoretical study[9,10,11,12]. Its experimental demonstration is made difficult by quasiparticle dissipation due to gapless excitations in nanowires or in vortex cores. This difficulty might be overcome by using certain strongly disordered superconductors in the vicinity of the superconductor-insulator transition (SIT). Here we report the first direct observation of the CQPS in a strongly disordered indium-oxide ($InO_x$) superconducting wire inserted in a loop, which is manifested by the superposition of the quantum states with different number of fluxes[13]. Similarly to the Josephson effect, our observation is expected to lead to novel applications in superconducting electronics and quantum metrology[1,10,11].**

Phase slips across superconducting wires[14] lead to non-zero resistance[2,15]; they also may lead to qubit dephasing due to charge noise[16] in the chain of Josephson junctions[17]. Resistance measurements are dissipative, so its saturation at low temperatures cannot be interpreted as the evidence for coherent QPS, however a blockade of supercurrent might indicate CQPS as demonstrated in ref. 8. Mooij and Harmans have suggested[13] a system decoupled from the environment: a superconducting loop in which phase slips change the number of quantized fluxes[18] resulting in their quantum superposition, therefore, exhibiting CQPS[9]. This letter reports a successful implementation of this idea using superconducting loops made of highly disordered $InO_x$.

We begin with considering a superconducting loop of an effective area $S$ with a high kinetic inductance $L_k$ containing a narrow segment (nanowire) with the finite rate of QPS. The external magnetic field $B_{ext}$ perpendicular to the loop induces the flux $\Phi_{ext} = B_{ext}S$. The states of the loop are described by the phase winding number, $N$, their energies are $E_N = (\Phi_{ext} - N\Phi_0)^2 / 2L_k$ (Fig. 1a). The energy difference between adjacent states |N+1⟩ and |N⟩ is $E_{N+1} - E_N = 2I_p\delta\Phi$, where $\delta\Phi = \Phi_{ext} - (N+1/2)\Phi_0$ and $I_p = \Phi_0/2L_k$ is the loop persistent current. At $\delta\Phi = 0$, the two states are degenerate. The QPS process, characterized by the amplitude $E_S$, couples the flux states resulting in the Hamiltonian

$$H = -\frac{1}{2}E_S(|N+1\rangle\langle N| + |N\rangle\langle N+1|) + E_N|N\rangle\langle N|, \qquad (1)$$

which is dual to the Hamiltonian of a superconducting island connected to a reservoir through a Josephson junction. In the latter, $L_k$ is replaced by capacitance, $E_S$ by the Josephson energy and $N$ is the number of the Cooper-pairs on the island[13]. The energy splitting between the ground and excited states $|g\rangle = \sin\frac{\theta}{2}|N\rangle + \cos\frac{\theta}{2}|N+1\rangle$ and $|e\rangle = \cos\frac{\theta}{2}|N\rangle - \sin\frac{\theta}{2}|N+1\rangle$ is $\Delta E = \sqrt{(2I_p\delta\Phi)^2 + E_S^2}$, where the mixing angle $\theta = \arctan[E_S/(2I_p\delta\Phi)]$.

To detect the CQPS, the loop is coupled to the coplanar line (resonator) via mutual inductance $M$[19]. In a rotating wave approximation, the effective Hamiltonian of the system resonantly driven by a classical microwave field with current amplitude $I_{MW}\cos(\Delta E t/\hbar)$, is $H_{RW} = \frac{\hbar\Omega}{2}(|e\rangle\langle g| + |g\rangle\langle e|)$, where $\hbar\Omega = MI_pI_{MW}\frac{E_S}{\Delta E}$. Note that transition between the two states can be induced only when $E_S \neq 0$, so observation of the spectroscopy peak constitutes a direct evidence of CQPS.

Now we provide a theoretical background for material choice. A CQPS is a result of quantum fluctuations of the order parameter. Generally, the impact of fluctuations is characterized by the Ginzburg parameter, $G_i$, which is essentially the inverse number of

Cooper pairs in a volume $\xi^3$: $G_i = (\xi^3 \nu \Delta)^{-1}$, where $\nu$ is the electron density of states, $\Delta$ is the superconducting gap and $\xi$ is the coherence length. Even in disordered bulk superconductors characterized by $k_F l \sim 1$, $G_i$ is small: $G_i \sim (k_F \xi)^{-1} \ll 1$ ($k_F$ is the Fermi wavevector and $l$ is the mean free path). Strong fluctuations require materials with an even higher degree of disorder, in which electrons are localized. The superconductivity is suppressed if the localization length, $\xi_{loc}$, becomes shorter than $\xi$. Following this reasoning we expect that the fluctuations are maximal ($G_i \sim 1$), when $\xi_{loc} \sim \xi$, that is close to SIT. Although Bardeen-Cooper-Schrieffer (BCS) theory[2] fails to describe SIT, it provides qualitatively correct explanation for behaviour of materials with $\xi_{loc} \sim \xi$. The fluctuations are enhanced in narrow wires with a small number of the effective conductive channels, $N_{ch} = R_K / R_\xi$ ($R_\xi$ is the resistance of the wire of length $\xi$ and $R_K = h/e^2$). The phase-slip amplitude decreases with $N_{ch}$ as $E_S \sim \exp(-aN_{ch})$[20,21] (here $a$ is the numerical parameter of order of 1, our data are consistent with $a \approx 0.3$ for $R_\xi$ = 1 kΩ and $\xi$ = 10 nm. (Supplementary Information)). Thus, $E_S$ is expected to be sizeable in superconducting strongly disordered quasi-1D wires. However, a high degree of disorder may also enhance Coulomb repulsion between electrons and turn the superconductor into a dissipative normal metal[22]. The ideal materials for observation of CQPS should therefore form localized preformed pairs even before they become superconducting[23,24,25]. Such behaviour has been observed in amorphous $InO_x$ and TiN thin films[26,27].

Our loops (Fig. 1**b**) are fabricated from a 35 nm thick superconducting $InO_x$ film with $\xi \approx 10 - 30$ nm[28], using electron-beam lithography. The uncertainty in definition of $\xi$ is due to poor applicability of BCS theory (Supplementary Information). The loops consist of wide parts and narrow segments of about 40 nm wide and 400 nm long (Fig. 1**c**). The wire is reasonably homogeneous, having variations of width less than 10 nm.

Identical wires and films, fabricated by the same process, have been characterized by dc transport measurements. The film exhibits superconducting transition at $T_c = 2.7$ K. The sheet resistance $R_{SQ}$ of the wide films, measured slightly above $T_c$ is 1.7 k$\Omega$, which gives rough (BCS) estimate of the sheet inductance $L_{SQ} \approx \hbar R_{SQ} / \pi \Delta \approx 0.7$ nH[13] ($\Delta \approx 0.5$ meV[26]) for the wide sections of the loops.

In order to measure the loops, we incorporate them in a step-impedance coplanar resonator – a strip InO$_x$ line of length $L = 1.5$ mm and width $W = 3$ μm, galvanically connected to two $Z_0 = 50\Omega$ gold coplanar lines at the ends (Fig. 1**d**). We estimate the effective wave impedance of the line to be $Z_1 = \sqrt{l_1 / c_1} \approx 1.6$ k$\Omega$ (where the specific inductance $l_1 = L_{SQ}/W \approx 2.3 \times 10^{-4}$ H/m and specific capacitance $c_1 \approx 0.85 \times 10^{-10}$ F/m). Because $Z_1 \gg Z_0$, the strong mismatch results in standing wave formation with maximal current at the boundaries: $I(x) = I_m \cos(\pi m x / L)$, where $x$ is the coordinate along the resonator ($0 \leq x \leq L$). The resonance frequency of the *m*-th mode $f_m$ ($m = 1,2,3...$) is given by $f_m \approx mv/2L = m \times 2.4$ GHz, where $v = (l_1 c_1)^{-1/2} \approx 7.2 \times 10^6$ m/s is the group velocity in the resonator. The energy decay rate in such a resonator is $\kappa \approx (4Z_0 / Z_1)(v/L)$, which limits the power peak width to $\kappa / 2\pi \approx 0.1$ GHz.

Now, we present our main results. Figure 2**a** shows the power transmission coefficient $|t|^2$ through the resonator at a temperature of 40 mK. The peaks correspond to $m = 3,4,5$ with resonance frequencies $f_3$ = 6.65 GHz, $f_4$ = 9.08 GHz and $f_5$ = 11.00 GHz (close to our estimates above). The actual peak widths (FWHM) are approximately 250 MHz ($=\kappa/2\pi$), which is twice as large as that expected from the loading loss. This is probably due to an extra loss in the gold ground plane films. Our loops located in the middle of the resonator ($x = L/2$) are coupled only to the odd modes $m$, for which the current defined by $I_{MW} = I_m \cos(\pi n/2)$ is non-zero.

To detect the superposition of flux states we measure transmission $t$ through the resonator at $f_m$ versus $B_{ext}$. The transmission does not depend on $B_{ext}$ at the 3$^{rd}$ and 5$^{th}$ peaks. However, at the 4$^{th}$ mode peak, $t$ exhibits well pronounced periodic structure: sharp negative dips in the amplitude $|t|$ together with phase $\arg(t)$ rotation, as shown in Fig. 2**b**. The period $\Delta B = 0.061$ mT corresponds with a high accuracy to one flux quantum $\Phi_0$ through the area $S \approx 32$ $\mu m^2$ of the loop shown in Fig. 1**b** with a 40 nm wide wire. Note that there is a small (~ 5%) uncertainty in the loop area definition due to the finite width of the lines. Each resonator contains five loops with the nanowire widths 40, 60, 80, 100 and 120 nm. Additionally to the wire widths, the loops slightly differ by their areas (by 10 – 12% from one to another). No CQPS-related signals are found from the samples with the nanowire segments wider than 40 nm, indicating that in these loops $E_S/h < 1$ GHz in a good agreement with our estimates (Supplementary Information) predicting suppression of $E_S$ by more than 10 times with increase of the nanowire width by 20 nm.

Although the coupling of our loop to the resonator ($g/h \sim 10$ MHz) is weak ($g \ll \hbar\kappa$), we were able to perform spectroscopy measurements by monitoring resonator transmission[29,30], while the frequency $f_{probe}$ of an additional probe microwave tone and $B_{ext}$ are scanned. The transmission phase plot shows the resonance excitation of the two-level system (Fig. 2**c**). The obtained structure is distorted by periodic resonances at $f_m$ (seen as red horizontal features) and the resulting picture is plotted after filtering out the resonances. The green-blue line corresponds to the expected energy splitting, which is well fitted by $\Delta E = \sqrt{(2I_p \delta\Phi)^2 + E_S^2}$ (the dashed line) with the fitting parameters $E_S/h = 4.9$ GHz and $I_p = 24$ nA, corresponding to $L_k \approx 42$ nH.

The exact splitting at the degeneracy point $\delta\Phi = 0$ happens to nearly coincide with $f_2 \approx 4.8$ GHz, therefore distorting the line shape. In Fig. 2**d**, we show the resonant peak at $\Phi_{ext} \approx 0.52\Phi_0$, slightly away of the degeneracy point, where the resonant frequency $\Delta E/h \approx 5.6$ GHz is between $f_1$ and $f_2$. The spectroscopy peak is well fitted by a Gaussian $A\exp\left[-\frac{1}{2}\left(\frac{f - \Delta E/h}{\Delta f}\right)^2\right]$ with $\Delta f = 260$ MHz. This demonstrates coherent coupling between the flux states in the loop. An interesting question, which requires further study, is the mechanism of decoherence in this type of systems. At this stage, we can only conjecture that the shape suggests peak broadening due to a low frequency Gaussian noise, rather than relaxation.

Below we discuss the relation between the measured and expected properties of the system. The total loop resistance (at $T > T_c$), $R$, is the sum of the resistances of the wide section estimated from $R_{SQ}$ and the nanowire resistance $R_W \approx 30$ kΩ, deduced from the dc measurements of identical samples. We estimate $R \approx 55$ kΩ. As one might expect the effective sheet resistance of the narrow wires is higher than $R_{SQ}$ (in our case by a factor of 1.8). BCS estimate of the inductance of the whole loop gives $L_k \approx 23$ nH. Close to the SIT, $R$ continuously rises with decreasing $T$ before developing superconductivity whereas $L_k$ diverges at SIT. Arguing by continuity, BCS theory is expected to underestimate $L_k$ in the strongly disordered superconductors (Supplementary Information). For our films, we find that $L_k$ is underestimated by the factor of 1.8 which is very reasonable for the superconductors close to SIT.

Another criterion of the wire quality is the value of its critical current. For a BCS superconductor the maximal supercurrent[13] is $I_c \approx 0.75\Delta/eR_\xi$ (Supplementary information). Taking into account that in the vicinity of SIT, $I_c$ is overestimated by at least the same factor of 1.8, (similar to $L_k$), we find $I_c \leq 280-90$ nA, for $\xi = 10-30$ nm. The measured value is $I_c = 100$ nA is in a full agreement with these expectations.

The quantitative estimate of the QPS amplitude is less accurate than for kinetic inductance or critical current due to its exponential dependence on the sample and material parameters. Furthermore in the BCS approximation, QPS amplitude explicitly depends on $\xi$ which is not well defined close to SIT. Rough estimates based on ref. 15 give values of ~5–20 GHz for $\xi = 10$ nm, our calculations, based on the measured value of the sheet kinetic inductance, yield $E_S$ ~10 GHz in agreement with the data (Supplementary information).

Below we present additional evidence for CQPS. First, although we have presented the data from a single sample, we observed very similar results in two other samples fabricated from the same $InO_x$ film. One of those samples was measured in the step-impedance resonator described above, and the other – in the open line configuration[19]. In all three samples, we found resonances with close values of persistent current and level splitting $E_S/h$ equal to 4.9 GHz, 5.8 GHz and 9.5 GHz, correspondingly. Note that the variation of $E_S$ is surprisingly low (given its exponential dependence on parameters), and is in agreement with expectations (Supplementary Information). This reproducibility is a strong argument against an alternative interpretation of the results based on unintentional formation of a rogue Josephson junction somewhere in the wire. Second, the energy difference between the two lowest states in the loop, similarly to the charge qubit (dual to our system with CQPS), asymptotically approach linear dependence $\Delta E = 2I_p \delta \Phi$ at $\Delta E \gg E_S$. In our experiments, we are able to trace the two-level system resonance up to $|\delta \Phi| \geq 0.5 \Phi_0$ and $\Delta E \approx 77$ GHz (Fig. 3). This linear dependence, without any observable splitting at $\Phi = 0$ and $\Phi_0$ (corresponding to the second order $4\pi$-process) indicates a linear inductance in our system, rather than non-linear one originating in a rogue Josephson junction. To support the statement, we present simulations of the spectroscopy lines (dotted cyan lines in Fig. 3) for an RF-SQUID qubit with a single Josephson junction (with $I_c = 100$ nA), capacitance and linear inductance, adjusted to provide the best fit of our spectroscopy data. The simulations show a significant rounding at $\Phi = 0$ and $\Phi_0$ due to the non-linear Josephson inductance in a qualitative disagreement with our results. These arguments allow us to exclude the rogue Josephson junction formed in our device, they cannot give us, however, a number of interfering phase slip locations in the wire, if, for example, the wire is not uniform.

Summarizing, we have demonstrated the coherent phase slip effect in a loop containing a nanowire fabricated from the highly disordered $InO_x$ superconducting film. Like the Josephson effect, the discovery opens prospects for many applications.

# References

[1]Mooij, J. E. & Nazarov, Yu. V. Superconducting nanowires as quantum phase-slip junctions. *Nature Physics* **2**, 169-172 (2006).

[2]Tinkham, M. "Introduction to superconductivity", McGraw-Hill, New York (1996).

[3]Avrein, D. V. Zorin A. B. & Likharev K. K. Bloch oscillations in small Josephson junction. Zh. Esp. Teor. Fiz. **88**, 407-412 (1984).

[4]Giordano, N. Evidence for macroscopic quantum tunnelling in one-dimensional superconductors. *Phys. Rev. Lett.* **61**, 2137-2140 (1988).

[5]Bezryadin, A., Lau, C. N. & Tinkham, M. Quantum suppression of superconductivity in ultrathin nanowires. *Nature* **404**, 971-974 (2000).

[6]Zgirski, M., Riikonen, K.-P., Touboltsev, V. & Arutyunov, K. Yu. Quantum fluctuations in ultranarrow superconducting aluminum nanowires. *Phys. Rev. B* **77**, 054508 (2008).

[7]Lehtinen, J. S., Sajavaara, T., Arutyunov K. Yu. & Vasiliev, A. Evidence of quantum phase slip effect in titanium nanowires. Preprint at (arXiv:1106.3852) (2011).

[8]Hongisto, T. T. & Zorin, A. B. Single charge transistor based on superconducting nanowire in high impedance environment. aXiv:1109.3634 (2011).

[9]Matveev, K. A. , Larkin, A. I. & Glazman, L. I. Persistent current in superconducting nanorings. *Phys. Rev. Lett.* **89**, 096802 (2002).

[10]Hriscu, A. M. & Nazarov, Yu. V. Model of a proposed superconducting phase slip oscillator: A method for obtaining few-photon nonlinearities. *Phys. Rev. Lett.* **106**, 077004 (2011).

[11]Hriscu, A. M. & Nazarov, Yu. V. Coulomb blockade due to quantum phase-slips illustrated with devices. *Phys. Rev.B* **83**, 174511 (2011).

[12]Vanevic, M. & Nazarov, Yu. V. Quantum phase slips in superconducting wires with weak links. arXiv:1108.3553 (2011).

[13]Mooij, J. E. & Harmans, C. J. P. M. Phase-slip flux qubits, *New Journal of Physics* **7**, 219 (2005).

[14]Little, W. A., Decay of persistent current in small superconductors. *Phys. Rev.* **156**, 396-403 (1966).

[15]Arutyunov, K. Yu., Golubev, D. S. & Zaikin, A. D. Superconductivity in one dimension. *Phys. Rep*. **464,** 1-70 (2008).

[16]Manucharyan, V. E., Masluk N. A., Kamal A., Koch, J., Glazman. L. I. & Devoret, M. H. *Phys. Rev. B* **85**, 024521 (2012).

[17]Pop, I. M., Douçot, B., Ioffe, L., Protopopov, I., Lecocq, F., Matei, I., Buisson, O. & Guichard, W., Experimental demonstration of Aharonov-Casher interference in a Josephson junction circuit. Preprint at ⟨http://arxiv.org/abs/1104.3999⟩ (2011).

[18]Hongisto, T. T. & Arutyunov, K. Yu. Suppression of diamagnetism in superconducting nanorings by quantum fluctuations.arXiv:0905.3464 (2011).

[19]Astafiev, O., Zagoskin, A. M., Abdumalikov, A. A., Pashkin, Yu. A., Yamamoto, T., Inomata, K., Nakamura, Y. & Tsai, J. S. Resonance fluorescence of a single artificial atom. *Science* **327**, 840-843 (2010).

[20]Zaikin, A. D., Golubev, D. S., van Otterlo, A. & Zimanyi, G. T. Quantum phase slips and transport in ultrathin superconducting wires. *Phys. Rev. Lett*. **78**, 1552-1555 (1997).

[21]Golubev, D. S. & Zaikin, A. D. Quantum tunneling of the order parameter in superconducting nanowires.*Phys. Rev. B* **64**, 014504 (2001).

[22]Finkelstein, A. M. Suppression of superconductivity in homogeneously disordered systems, *Physica B* **197**, 636-648 (1994).

[23]Feigel'man, M.V., Ioffe, L. B., Kravtsov, V. E. & Cuevas, E. Fractal superconductivity near localization threshold. *Annals of Physic s* **325**, 1390-1478 (2010).

[24]Feigel'man, M. V., Ioffe, L. B., Kravtsov, V. E. & Yuzbashyan E. A. Eigenfunction fractality and pseudogap state near the superconductor-insulator transition. *Phys. Rev. Lett.*, **98**, 027001 (2007).

[25]Feigel'man, M. V., Ioffe L. B. & Mezard, M. Superconductor-insulator transition and energy localization. *Phys. Rev. B* **82** 184534 (2010).

[26]Sacepe, B., Dubouchet, T., Chapelier, C., Sanquer, M., Ovadia, M., Shahar, D., Feigel'man, M. & Ioffe. L. Localization of preformed cooper pairs in disordered superconductors. *Nature Physics* **7**,239-244 (2011).

[27]Sacepe, B., Chapelier, C., Baturina, T. I., Vinokur, V. M., Baklanov, M. R. & Sanquer. M. Pseudogap in a thin film of a conventional superconductor. *Nature Communications* **1**, 140 (2010).

[28] Johansson, A., Sambandamurthy, G., Shahar, D., Jacobson, N. & Tenne, R. Nanowire acting as superconducting quantum device. *Phys. Rev. Lett.* **95** 116805 (2005).

[29] Wallraff, A., Schuster, D. I., Blais, A., Frunzio, L., Majer, J., Girvin S. M. & Schoelkopf, R. J. Approaching unit visibility for control of a superconducting qubit with dispersive readout. *Phys. Rev. Lett.* **95**, 060501 (2005).

[30] Abdumalikov, A. A., Astafiev, O. V., Nakamura, Y., Pashkin, Yu. A., & Tsai, J. S. Vacuum Rabi splitting due to strong coupling of a flux qubit and a coplanar-waveguide resonator, *Phys. Rev. B* **78**, 180502 (2008).

**Acknowledgments** We are grateful to M. Feigel'man, J. Mooij and Y. Nazarov for useful discussions. This work was supported by Funding Program for World-Leading Innovative R&D on Science and Technology (FIRST), MEXT KAKENHI "Quantum Cybernetics", Ministry of Science and Education of Russian Federation grant 2010-1.5-508-005-037. L. Ioffe was supported by ARO W911NF-09-1-0395, DARPA HR0011-09-1- 0009 and NIRT ECS-0608842. D. Shahar and O. Cohen were supported by Minerva Fund.

**Author Contributions** O.V.A. planned the experiment, designed and fabricated the samples, performed measurements and data analysis. L.B.I. came up with the decisive idea of using materials close to SIT and provided theoretical support of the work. S.K. fabricated the sample and contributed to the data understanding. Yu.A.P. participated in the discussions of the experiment. K.Yu.A. triggered the research direction and suggested to realize the phase-slip qubit. D.S. and O.C. fabricated InO films. J.S.Tsai discussed the data and provided the major support of the work within FIRST and KAKENHI projects. O.V.A. wrote the manuscript with feedback of all authors with significant contribution of L.B.I. and K.Yu.A.

**Competing Interests** The authors declare that they have no competing financial interests.

**Correspondence**    Correspondence and requests for materials should be addressed to O. V. Astafiev (astf@zb.nec.com.jp).

**Figure captions**

**Figure 1** | The device. **a**, Energies of the loop versus external flux $\Phi_{ext}$. The degeneracy between states with $N$ and $N+1$ of $\Phi_0$ is lifted due to the phase-slip energy $E_S$, when $\Phi_{ext} = (N + 1/2) \Phi_0$. **b** InO$_x$ loop with the narrow wire on the right side is attached to the resonator (straight line) at the bottom. **c**, False colour SEM image of the narrow InO$_x$ segment. **d**, The step-impedance resonator consisting of the 3 μm wide InO$_x$ strip with wave impedance $Z_1 \approx 1600$ Ω galvanically coupled to the gold $Z_0 = 50$ Ω coplanar line. The boundaries of the resonator are defined by the strong impedance mismatch ($Z_1 >> Z_0$)

**Figure 2** | Experimental data. **a,** Power transmission through the resonator measured within the bandwidth of our experimental setup. Transmission power coefficient $|t|^2$ peaks correspond to the resonator modes with numbers $m$ indicated for each peak. **b,** Transmission through the resonator as function of external magnetic field $B_{ext}$ at $m = 4$ ($f_4 = 9.08$ GHz). The periodic structure in amplitude $|t|$ and phase arg($t$) corresponds to the points where the lowest level energy gap $\Delta E/h$ matches $f_4$. The period $\Delta B = 0.061$ mT (= $\Phi_0/S$) indicates that the response comes from the loop, shown in Fig. 1**b**, with the effective loop area $S = 32$ μm$^2$. **c**, The two-level spectroscopy line obtained in two-tone measurements. The phase of transmission $t$ through the resonator at $f_4$ is monitored, while the other tone with frequency $f_{probe}$ from an additional

microwave generator is swept in a range from 2 to 15 GHz at different $B_{ext}$. The picture is filtered out to eliminate the contribution of other resonances ($2 \leq m \leq 6$), visible as horizontal red features. The dashed line is the fit to the energy splitting with $\Delta E/h$ = 4.9 GHz, $I_p$ = 24 nA. **d**, The resonant dip is measured at $\Phi/\Phi_0 = 0.52$. The red curve is the Gaussian fit.

**Figure 3** | Spectroscopy of the system in a wide flux and frequency ranges. The spectroscopy response is obtained by measuring variation of the transmission amplitude or phase through the resonator at $f_4$ = 9.08 GHz, while the frequency of the additional microwave tone $f_{probe}$ is swept. Depending on the range, the two-level system excitation frequency $\Delta E/h$ is derived from the direct (single-photon) excitation $\Delta E/h = f_{probe}$ (blue dots), two-photon process $\Delta E/h = f_{probe} + f_4$ (green dots), three-photon process $\Delta E/h = 2 f_{probe} + f_4$ (red dots). With these methods, using $f_{probe} \leq 35$ GHz, we could trace $\Delta E/h$ up to about 77 GHz. The dashed black line is the calculated energy splitting with $E_S$ = 4.9 GHz and $I_p$ = 24 nA. Perfect agreement of the experimental data with the calculated energy bands supports our interpretation: the quantum states of the system correspond to the superposition of the two adjacent flux states induced by CQPS. The rare scattered points result from the noise, resonator resonances and higher order excitation processes. For comparison the additional dotted cyan lines show the expected behaviour of a RF-SQUID qubit with $E_J/h$ = 50 GHz ($I_c$ = 100 nA), $C$ = 1.1 fF and $L$ = 38 nHn, which qualitatively disagrees with our data at the degeneracy points $\Phi_{ext} = 0$ and $\Phi_{ext} = \Phi_0$.

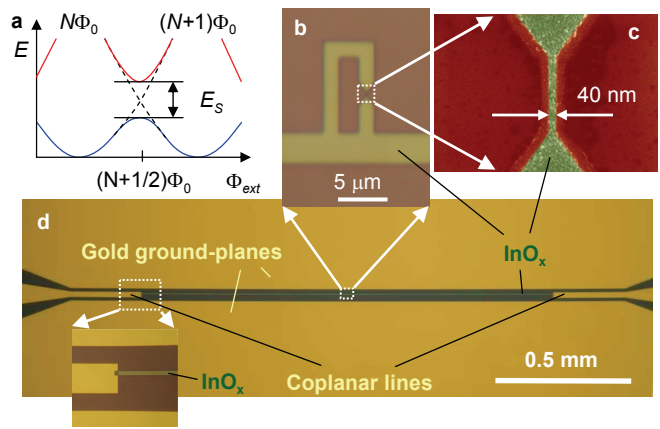

Figure 1

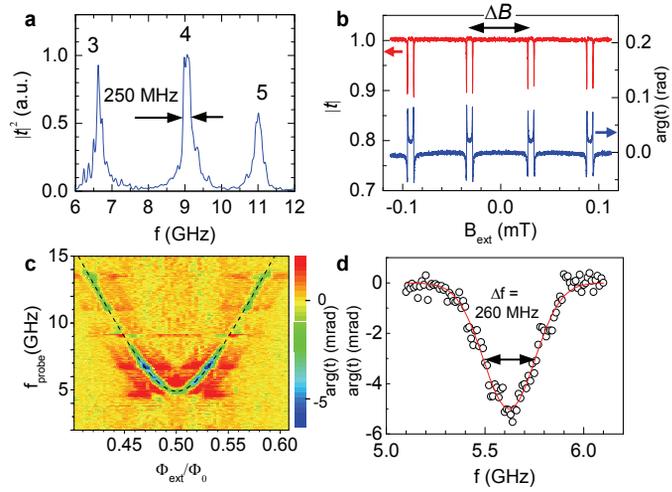

Figure 2

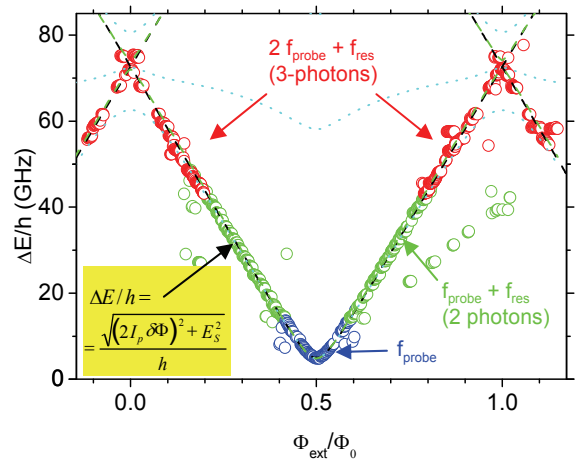

Figure 3

## Supplementary Information

**Film uniformity.** We begin by summarizing the results of previous studies that prove the amorphous nature of our films and the absence of a granular structure. We first note that our InO$_x$ films are identical to the ones studied extensively in a number of reports (see, e. g. refs. 26 31, 32, 33, 34 and ref. 35 for review) that focus on SIT: they were obtained in exactly the same conditions and in the same fabrication facility as most of the films from the papers of refs. 26, 32. All these films are prepared by electron-beam evaporation of high-purity In$_2$O$_3$ onto SiO$_2$ substrate. STM measurements at room temperatures and large voltages show that film roughness is less than 1 nm, for the typical two-dimensional map of the film surface, see e. g. the supplement to ref. 26. TEM of these films does not detect any crystal inclusions[31,32]. At low temperatures, STM shows very homogeneous density of states of electrons above the superconducting transition temperature and a very good superconducting gap everywhere on the surface of the film with no electron states below the gap at all points on the surface[26]. This excludes the granularity of the films and formation of insulating barriers that might lead to a Josephson network in the superconducting state. This also excludes the inhomogeneous state in which small droplets of superconductor are imbedded in a normal matrix. Furthermore, the coherence peaks appear simultaneously in all locations on the surface which implies that a superconducting condensate appears simultaneously in the whole sample. This is another proof of the absence of weakly connected grains of a good superconductor. Further discussion of the evidence against the granularity of these films can be found in review of ref. 35.

**Theoretical expectations for the phase-slip amplitude.** We now discuss the theoretical estimates of the phase-slip amplitude in different models. Moderately disordered films characterized by $k_F l > 1$ are described by BCS theory. In this case, a number of previous studies showed that the phase-slip rate is given by equation[15]

$$\gamma_{QPS} \approx \frac{\Delta}{\hbar} \frac{R_Q}{R_\xi} \frac{L}{\xi} \exp\left(-A \frac{R_Q}{R_\xi}\right), \tag{S1}$$

where $\Delta$ is the superconducting gap energy, $R_\xi$ is the resistance of the wire of length $\xi$, $L$ is the wire length, $R_Q = h/4e^2$ is the quantum resistance, and $a$ is the numerical coefficient. (Note that the factor $A$ in Eq. (S1) is translated into $a$ used in the main text by $a \approx A/4$, because $N_{ch} = R_K/R_\xi$ and $R_Q = R_K/4$.) In strongly disordered superconductors, this equation cannot be derived and has unclear meaning. In particular, the concept of the normal-state resistance that is well-defined (both theoretically and experimentally) in moderately disordered metals becomes ill-defined in the strongly disordered materials, where it is temperature dependent: In the absence of superconductivity the resistivity of these materials is expected to become infinite at zero temperature. Furthermore, strong fluctuations of electron properties associated with the disorder implies that the correlation length $\xi$ in these materials is also poorly defined and that it no longer satisfies the BCS relation $\xi^2 = \hbar l v_F / \pi \Delta$, where $v_F$ is the Fermi velocity. It is not surprising that different estimates of $\xi$ in these films give somewhat different values in the range 10 – 30 nm[28]. Using Eq. (S1) with $R_\xi = 1$ k$\Omega$, $\xi = 10$ nm and $\gamma_{QPS} \approx 2\pi \times 5$ GHz, we get the value of $A \approx 1.3$, and therefore $a \approx 0.3$, that matches the experimentally observed level splitting. The model[15,20,21], derived for moderately disordered superconductors, does not give a precise value of the numerical parameter $a$, except that it should be of the order of one.

The deviations from the predictions of the conventional BCS theory can be quantified by the comparison between resistance of the wire above the transition temperature $T_c$ and the kinetic inductance of the same wire at low temperatures. In the framework of the BCS theory they are related by $L_w = \hbar R_w / \pi \Delta$. This equation can be viewed as a consequence of the optical sum rule and a gap structure in BCS theory, therefore it remains valid even when the BCS relation $2\Delta = 3.52\ k_B T_c$ is violated by the strong coupling effects. In contrast, in a very disordered superconductor, the superfluid stiffness vanishes implying infinite $L_w$ while the resistance is measured at $T > T_c$ and the gap remains finite. Thus, one expects that as the superconductor gets very disordered the resistance taken for estimates and kinetic inductance should become larger than predicted by the BCS equation. For instance for our loop, the ratio of the measured inductance to the one expected in BCS theory is about 1.8, which confirms that these films are strongly disordered non-BCS superconductors but are still far from SIT where this ratio is expected to diverge.

In the regime of very strongly disordered superconductors, the properties are well described in the framework of the theoretical model of refs. 23, 24, 25. In this limit, the electrons form the localized pairs even in the absence of global superconducting coherence. The global coherence and superconductivity takes place only due to Cooper pair hopping from one localized state to another. In this model the phase-slip amplitude is expected to be[36]

$$E_S \approx \rho\sqrt{\frac{L}{W}}\exp\left(-\eta W\sqrt{\rho\nu d}\right), \hspace{2cm} (S2)$$

where $\rho$ is the phase stiffness per unit square, $h\nu \approx 4\times10^3$ GHz$^{-1}$μm$^{-3}$ is the electron density of states in the units convenient for this computations, $d$ is the film thickness, $W$ is width and dimensionless parameter $\eta \approx 1.0$. In two dimensional films, the phase stiffness defines the 2D current density phase gradient: $I = 2e\rho\partial\varphi/\partial x$ which has the dimension of energy and is directly related to the inductance per square: $\rho = \hbar/[(2e)^2 L_{SQ}]$. The directly measured value of the inductance per square $L_{SQ} \approx 2.2$ nH translates into $\rho/h \approx 70$ GHz: Although Eq. (S2) has very similar nature to Eq. (S1), Eq. (S2) has a number of advantages: it contains only experimentally measurable quantities, namely, superfluid stiffness and density of states and it can be directly verified by numerical simulations[36] in the strongly disordered regime. Using Eq. (S2), we estimate the phase amplitude to be $E_S \approx 5$ GHz for the wires of width $W \approx 40$ nm, remarkably close to the observation given the uncertainty in the value of the density of states and the approximations involved in the theoretical analysis. Furthermore, the numerical simulations[36] show that the phase slip amplitude varies by a factor of 2–3 from one disorder realization to another, also in agreement with the data. This agreement between the theory and the experiment confirms that the observed phase slip is a property of uniformly strongly disordered superconductors.

**Estimate of the critical current.** The maximal supercurrent in BCS approximation can be estimated using the equation

$$I_c \approx 0.75 \Delta / e R_\xi, \qquad (S3)$$

which is essentially the equation given in ref. 13. Substituting the wire resistance $R_W = 30$ k$\Omega$, and taking into account the same suppression factor as for inductance evaluation (~ 1.8), we obtain $I_c = 280 - 90$ nA for $\xi = 10 - 30$ nm.

Alternatively, the maximal critical current can be estimated using only the measurable quantities: the superconducting gap and density of states. The critical current of a thin film is related to the penetration depth $\lambda$ and the critical field $H_c$ of the bulk[37]

$$I_c = \frac{\sqrt{2}}{6\pi\sqrt{3}} \frac{c H_c}{\lambda}, \qquad (S4)$$

where $c$ is the speed of light. Expressing $\lambda$ and $H_c$ through $\Delta$ and $R_{SQ}$, one gets the critical current of the wire in BCS theory

$$I_c = \left(\frac{2\Delta}{3}\right)^{\frac{3}{2}} w \sqrt{\frac{\nu d}{\hbar R_{SQ}}}, \qquad (S5)$$

where $d$ is the wire thickness, $w$ is its width. Substituting $R_{SQ} = 3$ k$\Omega$ from the wire resistance, $\nu \approx 1.0$ eV$^{-1}$ nm$^{-3}$ and $\Delta \approx 0.5$ meV, we get $I_c \approx 400$ nA. Approaching SIT, $I_c$ vanishes, whereas Eqs. (S5) based on BCS assumptions gives a finite value. Similarly to superfluid stiffness, the critical current is suppressed in the vicinity of SIT. In fact, because it is a product the superfluid stiffness and a maximal gradient of the phase that does not destroy the superconductor, it is suppressed more than the superfluid stiffness. Thus, one expects that in superconductors close to

SIT, the critical current should be less than the one predicted by BCS equation at least by a factor of 1.8, obtained for the kinetic inductance correction, which results in $I_c \leq 200$ nA.

## References

[31] Ovadyahu, Some finite temperature aspects of the Anderson. transition Z. *J. Phys. C: Solid State Phys.* **19**, 5187-5213 (1986).

[32] Shahar, D. & Ovadyahu, Z. Superconductivity near the mobility edge. *Phys. Rev. B* 46, 10917-10922 (1992).

[33] Gantmakher, V. F., Golubkov, M. V., Dolgopolov, V. T., Tsydynzhapov, G. E. & Shahskin, A. A. Destruction of localized electron pairs above the magnetic-field-driven superconductor-insulator transition in amorphous In-O films. *JETP Letters* **68**, 363-369 (1998).

[34] Steiner M., Kapitulnik, A. Superconductivity in the insulating phase above the field-tuned superconductor-insulator transition in disordered indium oxide films. *Physica C* **422**, 16-26 (2005).

[35] Gantmakher, V.F. & Dolgopolov, V.T., Localized–delocalized electron quantum phase transitions. *UFN* **178**, 3–24 (2008).

[36] Feigelman, M. V., Ioffe, L. B. & Astafiev, O. V., Phase slip in strongly disordered superconducting wires, *unpublished*.

[37] Schmidt, V.V., Müller, P., Ustinov, A. V. & Grigorieva, I. V. "The Physics of Superconductors: Introduction to Fundamentals and Applications", Springer, (2010)